\documentclass[runningheads]{llncs}
\usepackage{multirow}
\usepackage{makecell}
\usepackage{color}
\usepackage{xcolor}
\usepackage{bbding}
\usepackage{amsmath}
\usepackage{graphicx}
\usepackage[marginal]{footmisc}

\usepackage[pagebackref=true,breaklinks=true,colorlinks,bookmarks=false]{hyperref}
\newcommand{\orcidlink}[1]{}
\usepackage[belowskip=6pt,aboveskip=5pt]{caption}
\title{Noise Level Adaptive Diffusion Model for Robust Reconstruction of Accelerated MRI}

\titlerunning{Noise Level Adaptive MRI Reconstruction}
\author{Shoujin Huang\inst{1}\, \orcidlink{0000-0001-6094-129X} \and
Guanxiong Luo\inst{2} \orcidlink{0000-0001-8005-4639} \and
Xi Wang\inst{3} \orcidlink{0000-0002-5218-2761} \and
Ziran Chen\inst{1} \orcidlink{0000-0003-1113-7697} \and \\
Yuwan Wang \inst{1} \orcidlink{0009-0002-0791-6989} \and 
Huaishui Yang\inst{1} \orcidlink{0009-0001-1656-9336}\and 
Pheng-Ann Heng\inst{3} \orcidlink{0000-0003-3055-5034}\and \\
Lingyan Zhang\inst{4} \and
Mengye Lyu\inst{1*} \orcidlink{0000-0001-5548-8136}}

\authorrunning{Shoujin Huang et al.}

\institute{Shenzhen Technology University, Shenzhen, China \and
Medical Center Göttingen University, Göttingen, Germany \and
The Chinese University of Hong Kong, Hong Kong, China \and
Longgang Central Hospital of Shenzhen, Shenzhen, China \\
*Corresponding authors: \email{lvmengye@sztu.edu.cn}}

\begin{document}

\maketitle              % typeset the header of the contribution
\vspace{-3em}
\footnote{{\footnotesize Shoujin Huang and Guanxiong Luo contribute equally to this work. Accepted by MICCAI (2024).}}

\begin{abstract}
In general, diffusion model-based MRI reconstruction methods incrementally remove artificially added noise while imposing data consistency to reconstruct the underlying images. However, real-world MRI acquisitions already contain inherent noise due to thermal fluctuations. This phenomenon is particularly notable when using ultra-fast, high-resolution imaging sequences for advanced research, or using low-field systems favored by low- and middle-income countries. These common scenarios can lead to sub-optimal performance or complete failure of existing diffusion model-based reconstruction techniques. Specifically, as the artificially added noise is gradually removed, the inherent MRI noise becomes increasingly pronounced, making the actual noise level inconsistent with the predefined denoising schedule and consequently inaccurate image reconstruction. To tackle this problem, we propose a posterior sampling strategy with a novel NoIse Level Adaptive Data Consistency (Nila-DC) operation. Extensive experiments are conducted on two public datasets and an in-house clinical dataset with field strength ranging from 0.3T to 3T, showing that our method surpasses the state-of-the-art MRI reconstruction methods, and is highly robust against various noise levels. The code for Nila is available at \url{https://github.com/Solor-pikachu/Nila}. 

\end{abstract}
\section{Introduction}\label{introduction}
Magnetic Resonance Imaging (MRI) is a crucial means for medical diagnostics, yet with major drawbacks of long acquisition times and high operational costs. To accelerate MRI scans, a common approach is using multiple coils to acquire the k-space data without obeying the Nyquist sampling theorem, and then to reconstruct the image with the incorporation of prior knowledge of it\cite{griswold2002generalized,pruessmann1999sense,sodickson1997simultaneous}, such as sparsity and total variation\cite{lustig2008compressed,block2007undersampled}. Deep learning models have been utilized to learn and apply prior knowledge inherently within existing databases, including both supervised\cite{sriram2020end,yiasemis2022recurrent,guo2023reconformer,huang2022swin} and generative paradigms\cite{vae_mri}. 

Recently, diffusion models have been applied to improve reconstruction, where the data consistency term of k-space is integrated as the guidance for the generative process\cite{jalal2021robust,fan2023survey,chung2022score,luo2023bayesian,gungor2023adaptive}, i.e, learned reverse process. In diffusion models\cite{ho2020denoising,song2020denoising}, there are: 1) the forward process created with Markov chains where the noise at different levels is added to images that represents the distribution of data; 2) the learned reverse process where the images of data distribution are generated  staring from a known distribution, i.e, Gaussian noise. 

\begin{figure*}[t]
    \centering
	\includegraphics[width=0.75
 \textwidth]{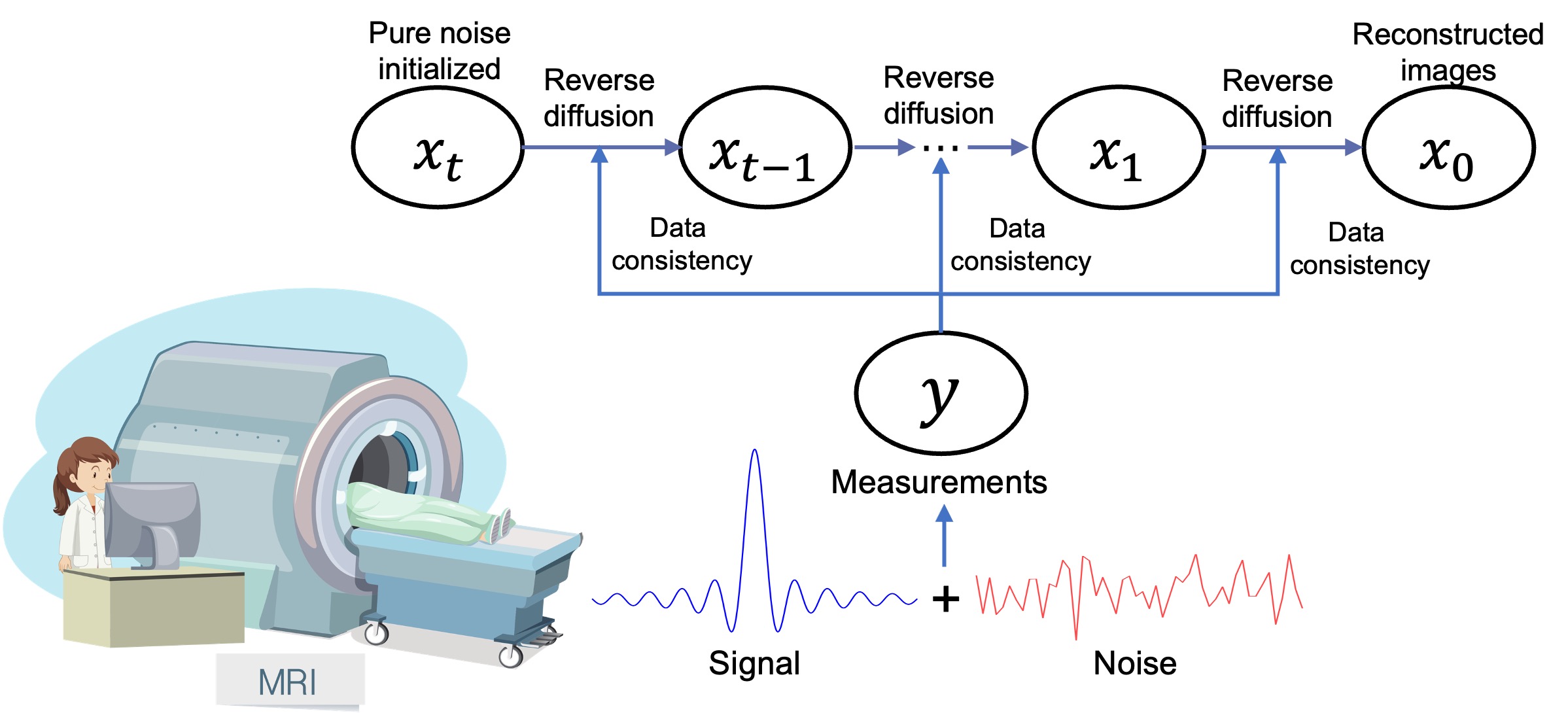}
	\caption{General illustration of diffusion model-based MRI reconstruction. Note that the MRI data (measurements) not only contain signals but also noise due to hardware and subject-related thermal fluctuations. This can interfere with the reverse diffusion process, where artificial Gaussian noise is used for image initialization and gradually removed with a pre-defined schedule.} \label{fig:problem}
\end{figure*}

However, hardware and subject-related thermal fluctuations cause the presence of noise in measured k-space. As illustrated in Fig. \ref{fig:problem},
this inherent noise propagates into the diffusion model-based reconstruction process through the data consistency term and interferes with the pre-defined noise schedule used for training the reverse process. This is an issue in the late stage of the reverse process when approaching the noise-free data distribution. Consequently, diffusion model-based reconstruction methods may experience sub-optimal performance or even failure when the MRI measurement noise is not negligible, such as in the case of low-field MRI\cite{marques2019low}, functional MRI (fMRI), and diffusion-weighted imaging (DWI)\cite{biswal1995functional}.

\begin{figure*}[t]
    \centering
	\includegraphics[width=0.95
 \textwidth]{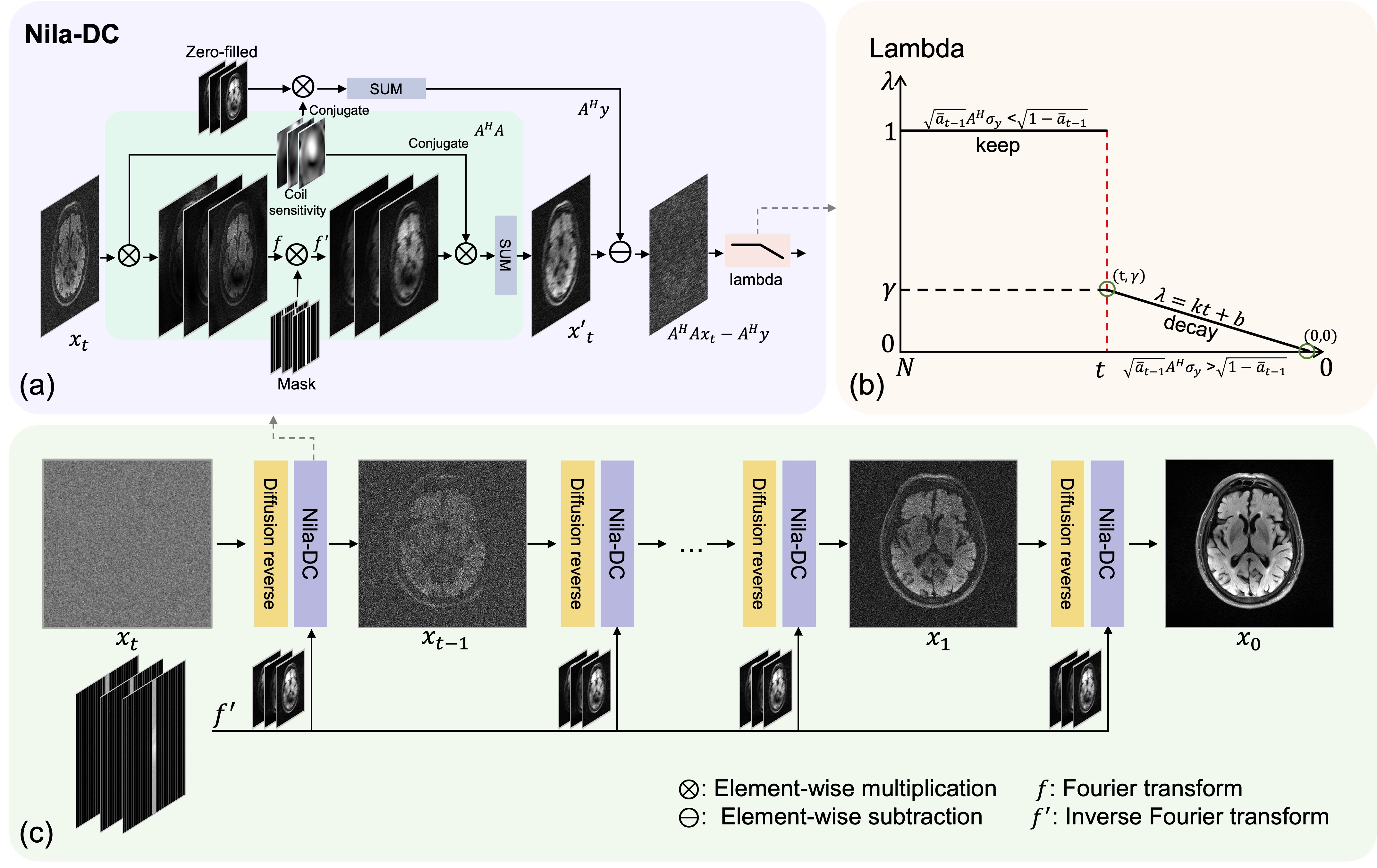}
	\caption{Overview of the proposed method. (a) The proposed data consistency (Nila-DC) operation. The computed gradient ($\mathcal{A}^H\mathcal{A}x_t-\mathcal{A}^Hy)$ can be noisy due to MRI noise in $y$ (c.f. Eqs.~\ref{eq7} and ~\ref{eqahay}) , and is therefore adjusted by a attenuation function (lambda). (b) The attenuation function (c.f. Eq.~\ref{eq:lambda}) used to rescale the DC gradient. $t$ is the index of the reverse step. (c) The image reconstruction process, where Gaussian noise initialized $x_{t}$ undergoes multi-step reverse diffusion process with the guidance from Nila-DC.} \label{fig:architecture}
\end{figure*}

In this work, a noise level adaptive data consistency (Nila-DC) operation is proposed for image reconstruction with a lambda rescale function (c.f.  Fig.~\ref{fig:architecture} (b)) that ensures robust guidance of k-space regardless of noise when using diffusion models. This approach is validated on multiple datasets of different field strengths, imaging sequences, and noise levels, including challenging scenarios such as accelerated DWI and low-field MRI. In the following sections, the disturbance of noise to the learned reverse process is formulated and experimental results using Nila-DC are presented.

\section{Methodology}\label{method}
Alike many diffusion model-based techniques, reconstruction is one of the samples from the posterior distribution which is formulated with two components: 1) the likelihood term to describe the data consistency; 2) the diffusion prior trained on image dataset\cite{jalal2021robust,gungor2023adaptive,luo2023bayesian,luo2023generative}. We observed that the different noise level in many acquisition scenarios affects the pre-defined noise schedule for training a diffusion model and consequently leads to poor performance. In the below, we present a noise level adaptive method to address this issue in the context of using the diffusion denoised probabilistic model (DDPM)  as a prior\cite{nichol2021improved,dhariwal2021diffusion}. % add citation

\subsubsection{Reconstruction as Bayesian Inversion.}
Image reconstruction is approached as a Bayesian problem, where the posterior $p(\mathbf{x}\mid\mathbf{y})$ is determined by the measured k-space $\mathbf{y}$ and a diffusion prior $p(\mathbf{x})$ is trained on an image database\cite{luo2023bayesian,luo2023generative}. Assuming the noise \textbf{$\eta$} normally distributed with zero mean and
covariance matrix $\sigma^2_{y} \mathbf{I}$, the
likelihood $p(\mathbf{y}\mid\mathbf{x})$ for observing the $\mathbf{y}$ 
determined by $\mathbf{y} = \mathcal{A} \mathbf{x}+\eta$
and given the image $\mathbf{x}$ is given by a complex normal distributions
\begin{align}
	p(\mathbf{y}\mid\mathbf{x}) & = \mathcal{CN}(\mathbf{y}; \mathcal{A}\mathbf{x}, \sigma^2_{y} \mathbf{I}) \nonumber\\
	& = (\sigma_{y}^2\pi)^{-N_p} \exp(\text{-}\|\sigma_{y}^{-1} \cdot (\mathbf{y} - \mathcal{A}\mathbf{x})\|_2^2)~,
	\label{eq:forw} 
\end{align}
The forward operator $\mathcal{A}$ is composed of coil sensitivity maps $\mathcal{S}$, a two-dimensional Fourier transform $\mathcal{F}$, and a k-space sampling operator $\mathcal{P}$. 
Using Bayes' formula,  we obtain each intermediate distribution in Fig.~\ref{fig:architecture}.
\begin{equation}
p\left(\mathbf{x}_t \mid \mathbf{y}\right) \propto p\left(\mathbf{x}_t\right) p\left(\mathbf{y} \mid \mathbf{x}_t\right)
\end{equation}
Starting with the initial Gaussian noise $q(\mathbf{x}_N)\sim \mathcal{CN}(0,\mathbf{I})$ at $t=N$, $p(\mathbf{x}_t)$ is obtained with transition kernels $\{p(\mathbf{x}_t | \mathbf{x}_{t+1})\}_{t\leq N}$ 
\begin{equation}
p(\mathbf{x}_t) \propto p(\mathbf{x}_t | \mathbf{x}_{t+1}) \cdot ... \cdot p(\mathbf{x}_{N-1} | \mathbf{x}_N) \cdot q(\mathbf{x}_N).
\end{equation}
With the learned transition kernel $p_\theta(\mathbf{x}_t \mid \mathbf{x}_{t+1})$ in denoising diffusion models, samples are simulated from $p_\theta(\mathbf{x}_t \mid \mathbf{y})$ using the reverse diffusion\cite{song2020denoising}:
\begin{equation}
    \label{eq4}
    \mathbf{x}_{t-1} = \sqrt{\overline{a}_{t-1}}\nabla_{\mathbf{x}_t} \log p_\theta(\mathbf{x}_t \mid \mathbf{y}) + \sqrt{1-\overline{a}_{t-1}} \mathbf{z}, \quad \mathbf{z} \sim \mathcal{CN}(0, \mathbf{I})~,
\end{equation}
where $\overline{a}_{t-1}:=\prod \limits_{s=1}^{t-1} {a}_{s}$, ${a}_{t-1}:=1-\beta_{t-1}$, $\beta_{t}$ following a pre-defined schedule $\{\beta_{0},\beta_{1}, ... , \beta_{T}\}$ used in the forward process\cite{ho2020denoising} , and  
\begin{equation}
    \label{eq5}
    \nabla_{\mathbf{x}_t} \log p_\theta(\mathbf{x}_t \mid \mathbf{y}) = \nabla_{\mathbf{x}_t} \log p(\mathbf{x}_t\mid\mathbf{x}_{t-1}) + \nabla_{\mathbf{x}_t} \log p(\mathbf{y} \mid \mathbf{x}_t)~.
\end{equation}
Further, we have
\begin{align}
    \label{eq6}
    \nabla_{\mathbf{x}_t}\log p(\mathbf{x}_t\mid\mathbf{x}_{t-1}) &=\frac{1}{\sqrt{\overline{a}_{t}}}(\mathbf{x}_{t}-\sqrt{1-\overline{a}_{t}}(\epsilon_{\theta}(\mathbf{x}_{t},t))) \\
    \label{eq7}
    \nabla_{\mathbf{x}_t} \log p(\mathbf{y} \mid \mathbf{x}_t) &= -(\mathcal{A}^{H}\mathcal{A}\mathbf{x}_t - \mathcal{A}^{H}\mathbf{y}) / \sigma_{y}^2
\end{align}
Eq.~\ref{eq6} represents the diffusion prior term, and Eq.~\ref{eq7} the data consistency term. Note that Eq.~\ref{eq7} can be derived without approximation errors\cite{chung2022diffusion,song2021solving} because of the MRI specific assumption in Eq.~\ref{eq:forw}. Initially, when $t$ is large, $\mathbf{x}_t$ has less structural information and higher noise level. When $t$ approaches 0, $\mathbf{x}_t$ is expected to recover structural information and have decreased noise level. However, it is important to note that $\mathbf{y}$ is noisy in nature.
\subsubsection{Noise Level Adaptive Data Consistency.}
During the computation of data consistency, the adjoint of the k-space represents the under-sampled k-space with additive noise. 
Then we have
\begin{equation}
    \label{eqahay}
    \mathcal{A}^H\mathbf{y} = \mathcal{A}^H\mathcal{A}\bar{\mathbf{x}}\ + \ \mathcal{A}^H\sigma_y, 
\end{equation}
where $\bar{\mathbf{x}}$ is the noise-free image and $\sigma_{y}$ is relative to noise level of images in the training dataset (note that $\sigma_{y}$ could be estimated via a quick calibration scan or from image background areas). This $\sigma_{y}$ propagates through the diffusion reverse process, adding extra noise $\sigma_{y}$ in the image $\mathbf{x}_{t}$. According to Eqs.~\ref{eq4},~\ref{eq5},~\ref{eq6},~\ref{eq7}, ~\ref{eqahay} and using gradient based sampling method, we have
\begin{align}
    \label{eq:noise}
    \mathbf{x}_{t-1}& =  \sqrt{\overline{a}_{t-1}}\left[\frac{1}{\sqrt{\overline{a}_{t}}}(\mathbf{x}_{t}-\sqrt{1-\overline{a}_{t}}\epsilon_{\theta}(\mathbf{x}_{t},t))-(\mathcal{A}^{H}\mathcal{A}\mathbf{x}_t -\mathcal{A}^H\mathcal{A}\bar{\mathbf{x}})\  \right]\nonumber\\
     &\phantom{=}+ \sqrt{\overline{a}_{t-1}}\mathcal{A}^H\sigma_y + \sqrt{1-\overline{a}_{t-1}} \mathbf{z}, \quad \mathbf{z} \sim \mathcal{CN}(0, \mathbf{I})~,
\end{align}
Although the introduction of noisy measurements by data consistency operation may not cause problems during the early stage of the reverse process, as the noise in $x_{t}$ is gradually removed, $\sigma_{y}$ from noisy measurement will become dominant at a certain stage, disturbing the pre-defined denoising schedule. Hence, we propose a new noise level adaptive data consistency operation to robustly utilize of guidance information from MRI data while minimizing the impact of MRI noise. From the pre-defined denoising schedule $\beta_{t} \ in \{\beta_{0},\beta_{1}, ... , \beta_{T} \}$, we know that the noise level at $\mathbf{x}_{t-1}$ is $\sqrt{1-\overline{a}_{t-1}}$; According to Eq.~\ref{eq:noise}, the additive noise propagating from the data consistency operation into the diffusion reverse process is $\sqrt{\overline{a}_{t-1}}\mathcal{A}^H\sigma_y$. To keep denoising schedule $\beta_{t}$ unchanged, we introduce a rescale function (Fig.~\ref{fig:architecture}(b)) to keep/decay the data consistency term. For simplicity, we use a linear attenuation function $\lambda_{t} = kt + b$, starting with $\gamma$ and linearly decaying to $0$ at the final moment:
\begin{align}
    \label{eq:lambda}
    \lambda_{t} = & \begin{cases}
1&, \sqrt{\overline{a}_{t-1}}\mathcal{A}^H\sigma_y<\sqrt{1-\overline{a}_{t-1}}\\ 
kt+b&, \sqrt{\overline{a}_{t-1}}\mathcal{A}^H\sigma_y>\sqrt{1-\overline{a}_{t-1}}
\end{cases},
\end{align}
$k$ and $b$ can be computed once $\sigma_y$ and $\gamma$ have been defined. In practice, step size is scaled by $\lambda_{t}$. Finally, each reverse step in the proposed MRI reconstruction is defined by:
\begin{align}
    \label{eq10}
    \mathbf{x}_{t-1}& = \sqrt{\overline{a}_{t-1}}\left[\frac{1}{\sqrt{\overline{a}_{t}}}(\mathbf{x}_{t}-\sqrt{1-\overline{a}_{t}}\epsilon_{\theta}(\mathbf{x}_{t},t))-\lambda_{t}(\mathcal{A}^{H}\mathcal{A}\mathbf{x}_t - (\mathcal{A}^H\mathcal{A}\bar{\mathbf{x}}\ + \ \mathcal{A}^H\sigma_y)) \right]\nonumber\\
     &\phantom{=}+ \sqrt{1-\overline{a}_{t-1}} \mathbf{z}, \quad \mathbf{z} \sim \mathcal{CN}(0, \mathbf{I})~.
\end{align}
\section{Experiments}\label{experiments}
\subsubsection{Datasets and Baselines.}
We evaluated our approach on three datasets:
1) The public fastMRI dataset\cite{zbontar2018fastmri}. From the official validation set, we randomly selected 20 samples for each contrast (T1, T1 post contrast, T2, and FLAIR) to form a test set of 80 samples. 2) The public low-field dataset M4Raw\cite{lyu2023m4raw}. Based on its v1.6 release, 25 individuals from the test set, each with T1, T2, and FLAIR data, were used to form another test set (75 samples in total). 3) An additional in-house clinical dataset. With written informed consent obtained, T1 and FLAIR data were collected from 35 patients with white matter lesions from a local hospital using a 3T MRI Siemens scanner. The proposed method Nila is compared with L1-wavelet SENSE reconstruction\cite{uecker2016bart}, diffusion model-based MRI reconstruction method CSGM\cite{jalal2021robust}, Spreco\cite{luo2023bayesian}, AdaDiff\cite{gungor2023adaptive}. Following\cite{zbontar2018fastmri}, we used  equidistant undersampling masks for all experiments because of ease of implementation on MRI machines. 

\subsubsection{Training and Evaluation Details.}
The multi-coil images from fastMRI\cite{zbontar2018fastmri} brain training set is coil-combined using ESPIRiT\cite{uecker2016bart} and then used to train our model, using code from guided diffusion\cite{dhariwal2021diffusion}. The last four noisy slices were removed from each volume, resulting in a total of {$\sim$}52k slice images for training. We used a learning rate of 0.0001, batch size of 8, and Adam optimizer. Note that this fastMRI trained model is also used to inference on the M4Raw and clinical datasets. For CSGM and Spreco, their official pre-trained models were used\cite{jalal2021robust,luo2023bayesian}, whereas the AdaDiff model was trained on the same dataset as Nila.

To explore the robustness of different methods, the above models were tested both with and without adding extra Gaussian noise to the fully sampled MRI data. Peak signal-to-noise ratio (PSNR) and structural similarity (SSIM) metrics were computed within the brain region, ignoring background reconstruction errors. Because low-field data is noisy, when evaluating on M4Raw, we set $\sigma_{y}=0.05$ for Nila and employed multiple repetition averaged images \cite{lyu2023m4raw} (6 repetitions for T1 and T2; 4 repetitions for FLAIR) as the reference to calculate the metrics.

\subsubsection{Quantitative Results.}
Even without adding extra noise, our algorithm consistently achieved highest PSNR and SSIM scores on all datasets (Table~\ref{tabel:sota table}). The differences between our method and others are statistically significant as shown in the supplementary materials. The second best method Spreco performed slightly better than CSGM and Adadiff, likely due to its early stopping mechanism. 

When the noise level was respectively controlled (Table~\ref{tabel:sota noisy table}), all algorithms had reasonable results at low noise level ($\sigma=0.025$). However, when noise increased, the reconstruction quality of L1-wavelet, CSGM, and Adadiff declined dramatically, sometimes even lower than Zero-filled reconstruction. In contrast, our algorithm remained resilient, achieving the highest scores for scenarios. 
\begin{table}
\caption{Quantitative assessment for the three test datasets at different acceleration factors. No extra noise was added. The highest scores are marked in red, second highest blue.}
\label{tabel:sota table}
\resizebox{1\textwidth}{!}{
\begin{tabular}{c|cc|cc|cc|cc|cc|cc}
\hline
\hline
\multicolumn{1}{c}{Dataset} & \multicolumn{4}{|c}{fastMRI} & \multicolumn{4}{|c}{Clinical} & \multicolumn{4}{|c}{M4Raw} \\
\hline
\multicolumn{1}{c}{Acceleration factor} & \multicolumn{2}{|c}{6$\times$} & \multicolumn{2}{|c}{8$\times$} & \multicolumn{2}{|c}{6$\times$} & \multicolumn{2}{|c}{8$\times$} & \multicolumn{2}{|c}{3$\times$} & \multicolumn{2}{|c}{4$\times$} \\
\hline
Metric & PSNR & SSIM & PSNR & SSIM & PSNR & SSIM & PSNR & SSIM & PSNR & SSIM & PSNR & SSIM \\
\hline
Zero-filled & 26.12 & 82.27 & 25.41 & 80.36 &  27.07 & 82.40 & 26.33 & 80.34 & 27.15 & 87.81 & 25.97 & 85.43 \\
L1-wavelet & 30.04 & 89.43 & 27.31 & 84.55 & 31.27 & 89.60 & 28.35 & 84.59 & 28.86 & 87.50 & 28.02 & 86.57\\
CSGM & \textcolor{blue}{35.04} & \textcolor{blue}{93.53} & \textcolor{blue}{32.11} & \textcolor{blue}{90.78} & \textcolor{blue}{34.67} & \textcolor{blue}{92.65} & \textcolor{blue}{31.57} & \textcolor{blue}{88.94} & 26.73 & 84.53 & 25.63 & 82.78 \\
Spreco & 32.24 & 90.70 & 28.65 & 85.28 & 32.27 & 89.84 & 28.77 & 84.07 & \textcolor{blue}{29.77} & \textcolor{blue}{90.34} & \textcolor{blue}{28.88} & \textcolor{blue}{88.68} \\
AdaDiff & 33.25 & 91.79 & 29.62 & 86.66 & 32.84 & 90.81 & 29.60 & 85.65 & 29.10 & 88.29 & 28.10 & 86.22 \\
Nila (ours) & \textcolor{red}{37.08} & \textcolor{red}{95.74} & \textcolor{red}{34.82} & \textcolor{red}{94.21} & \textcolor{red}{36.10} & \textcolor{red}{94.34} & \textcolor{red}{33.54} & \textcolor{red}{91.55} & \textcolor{red}{30.23} & \textcolor{red}{91.59} & \textcolor{red}{29.86} & \textcolor{red}{90.06} \\
\hline
\hline
\end{tabular}}
\end{table}
\begin{table}
\caption{Quantitative assessment under various noise levels at 6$\times$ acceleration factor. Note that the sigma values represent the added noise to full sampled data. The highest scores are marked in red, second highest blue.}
\label{tabel:sota noisy table}
\resizebox{1\textwidth}{!}{
\begin{tabular}{c|cc|cc|cc|cc|cc|cc}
\hline
\hline
\multicolumn{1}{c}{Dataset} & \multicolumn{6}{|c}{fastMRI} & \multicolumn{6}{|c}{Clinical} \\ 
\hline
noisy level & \multicolumn{2}{|c}{$\sigma=0.025$} & \multicolumn{2}{|c}{$\sigma=0.05$} & \multicolumn{2}{|c}{$\sigma=0.1$} & \multicolumn{2}{|c}{$\sigma=0.025$} & \multicolumn{2}{|c}{$\sigma=0.05$} & \multicolumn{2}{|c}{$\sigma=0.1$}  \\
\hline
Metric & PSNR & SSIM & PSNR & SSIM & PSNR & SSIM & PSNR & SSIM & PSNR & SSIM & PSNR & SSIM \\
\hline
Zero-filled & 26.06 & 81.40 &  25.90 & 79.22 & 25.32 & 74.08 & 27.01 & 81.73 & 26.84 & 79.96 & 26.23 & \textcolor{blue}{75.31}\\
L1-wavelet & 29.07 & 85.86 & 26.78 & 78.15 & 21.28 & 64.98 & 30.15 & 85.86 & 27.55 & 77.51 & 21.22 & 63.77 \\
CSGM & 31.15 & 87.64 & 25.08 & 76.16 & 19.47 & 61.78& \textcolor{blue}{31.53} & 87.16 & 25.93 & 75.73 & 19.64 & 61.52 \\
Spreco & \textcolor{blue}{31.31} & \textcolor{blue}{89.03} & \textcolor{blue}{29.62} & \textcolor{blue}{84.93} & \textcolor{blue}{25.89}  &  \textcolor{blue}{74.23}  &  31.32 & \textcolor{blue}{88.03} & \textcolor{blue}{29.58} & \textcolor{blue}{83.88} & \textcolor{blue}{26.41} & 74.10  \\
AdaDiff & 29.29 & 82.93 & 25.83 & 73.8 & 21.99 & 64.82 & 29.45 & 82.74 & 26.04 & 73.50 & 22.12 & 64.45 \\
Nila (ours) & \textcolor{red}{34.98} & \textcolor{red}{93.96} & \textcolor{red}{33.56} & \textcolor{red}{92.49} & \textcolor{red}{31.86} & \textcolor{red}{90.38} & \textcolor{red}{34.11} & \textcolor{red}{91.53} & \textcolor{red}{32.80} & \textcolor{red}{89.59} & \textcolor{red}{31.25} & \textcolor{red}{87.15} \\
\hline
\hline
\end{tabular}}
\end{table}
\begin{figure}
    \centering
    \includegraphics[width=1.0
     \textwidth]{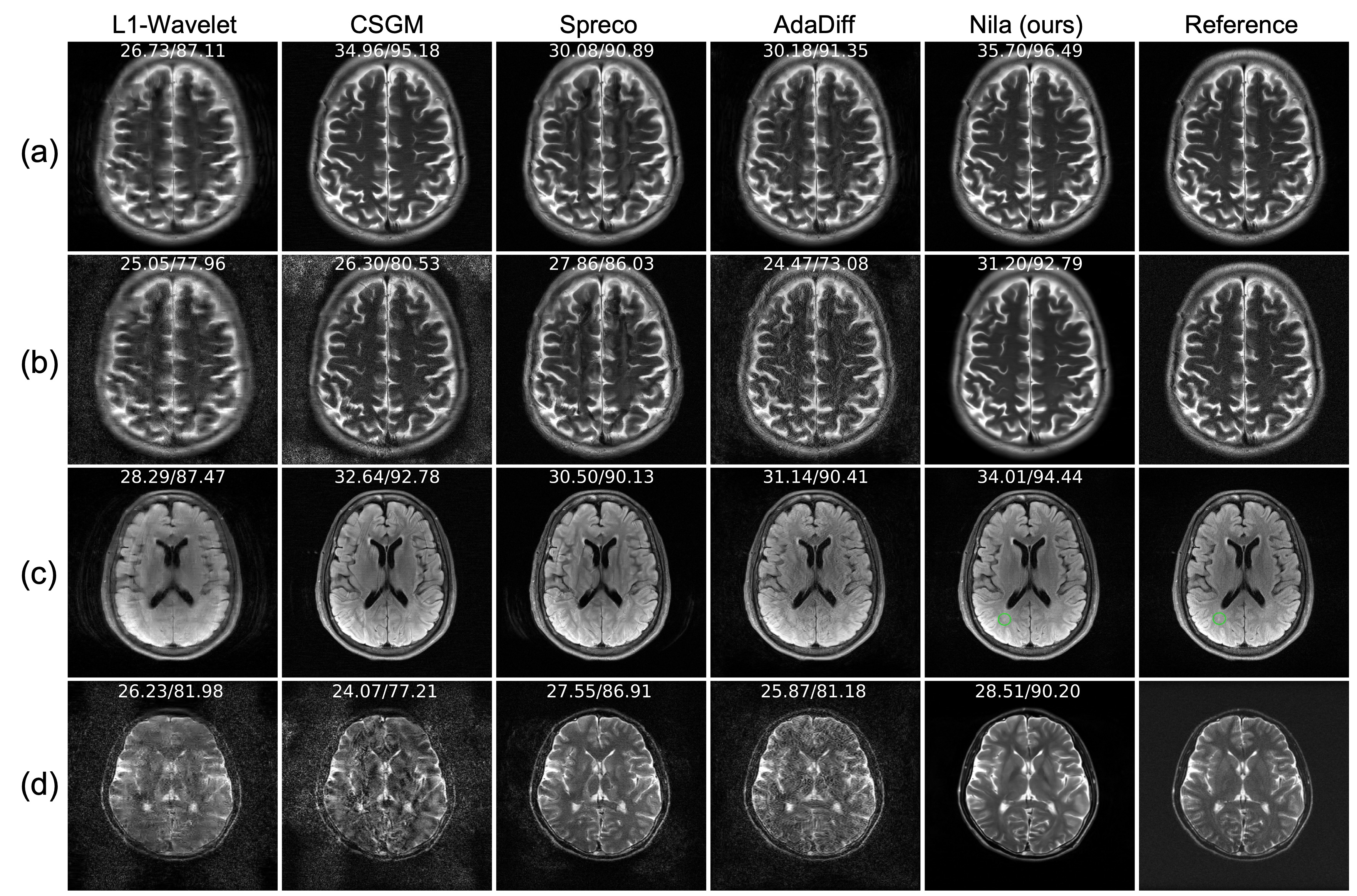}
    \caption{Typical reconstructed images. The white numbers indicate the PSNR/SSIM scores. (a) 6$\times$  acceleration on fastMRI. (b) 6$\times$  acceleration on fastMRI with added Gaussian noise. (c) 6$\times$  acceleration on the clinical dataset. Only the proposed method recovered the small white matte lesion as highlighted. (d) 4$\times$  acceleration on M4Raw. CSGM and AdaDiff failed to provide usable images.}
    \label{fig:vis}
    \end{figure}

\subsubsection{Qualitative Evaluation.}
Representative reconstruction results are presented in Fig.~\ref{fig:vis}. Consistent with the qualitative metrics, our algorithm had the highest reconstruction quality regardless of noise levels, clearly revealing the anatomical structures as well as small white matter lesions. 

\subsubsection{Real-World DWI Reconstruction.}
\begin{figure}[h]
\centering
\includegraphics[width=1.0 \textwidth]{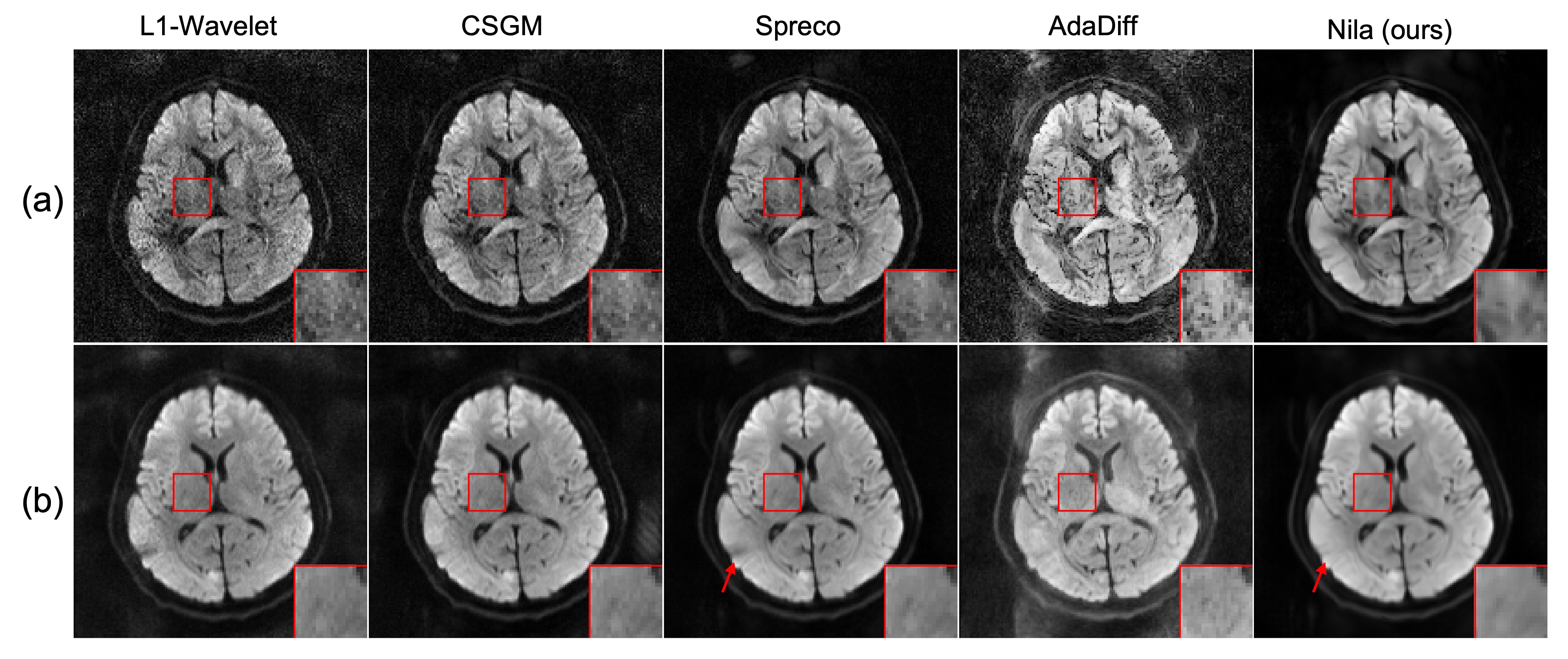}
\caption{Reconstruction of prospectively accelerated DWI data. (a) Reconstruction from a single repetition. (b) Averaged from 3 orthogonal diffusion weighting directions, each by 3 repetitions. $\sigma_{y}$ was set to 0.05 for Nila, as estimated from zero-filled reconstruction.}
\label{fig:dwi}
\end{figure}

Additionally, we experimented with prospectively undersampled 1.5T DWI data (3$\times$ acceleration, b-value = 800). Due to the nature of DWI, ground truth images are not possible to obtain here, but it is obvious that our method again achieved the highest SNR and the least reconstruction artifacts (Fig.~\ref{fig:dwi}). 

\begin{figure}
\centering
\includegraphics[width=1
 \textwidth]{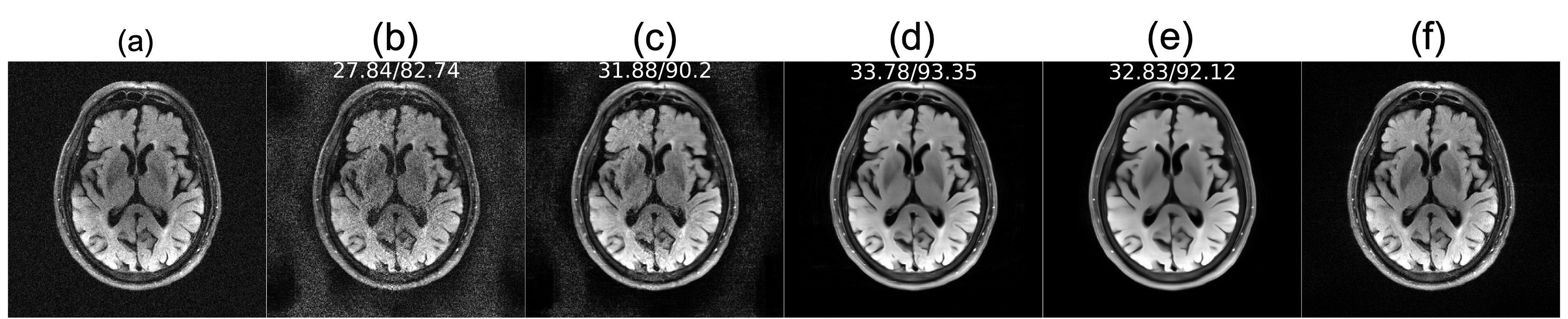}
\caption{Reconstruction of Nila with added noise $\sigma=0.05$ under different settings. (a) Fully sampled image with added $\sigma=0.05$ noise. It is for visualization of the noise level. (b - e) Nila reconstruction using $\sigma_{y}$ values of $0$, $0.025$, $0.05$, and $0.1$, respectively. (d) Fully sampled image without adding extra noise (clean reference). With $\sigma_y$ set to 0, Nila falls back to conventional DDPM-based MRI reconstruction like Eq.~\ref{eq:noise}. The white numbers indicate the PSNR/SSIM scores of the displayed images.}
\label{fig:abalation}
\end{figure}
\subsubsection{Ablation Study.}
The hyperparameter $\sigma_y$ in Nila affects its reconstruction quality. In particular, when $\sigma_y$ is set to 0, Eq.~\ref{eq:lambda} becomes an identity function without attenuation effect, and Nila-DC falls back to conventional data consistency operation. To examine the contribution of the proposed Nila-DC operation and the optimal settings of $\sigma_y$, we compared using different $\sigma_y$ on the clinical dataset at 6$\times$ acceleration with added Gaussian noise of $\sigma=0.05$. As shown in Fig.~\ref{fig:abalation}, setting $\sigma_y$ approximately equal to $\sigma$ resulted in the best reconstruction. With underestimated $\sigma_y$, the reconstruction yield amplified noise, whereas with overestimated $\sigma_y$, the reconstruction exhibited over-smoothing. In practice, $\sigma_{y}$ can be easily estimated from a quick calibration scan, e.g. using zero flip angle to acquire a few k-space lines, or background areas of zero-filled reconstruction.
\section{Conclusion}
We identify and address the issue that existing diffusion model-based reconstruction methods are sensitive to the MRI noise level by introducing a noise level adaptive data consistency operation for the reverse diffusion process, which permits robust guidance. The proposed method is comprehensively evaluated to demonstrate outstanding performance under various experimental conditions.
\subsubsection{Acknowledgments.}
This work was supported in part by the National Natural Science Foundation of China under Grant 62101348, the Shenzhen Higher Education Stable Support Program under Grant 20220716111838002, the Natural Science Foundation of Top Talent of Shenzhen Technology University under Grants 20200208, and the Research Grants Council of the Hong Kong Special Administrative Region, China (Project Reference Number: T45-401/22-N).

\subsubsection{Disclosure of Interests.}
The authors declare no competing interests.

% \newpage
\bibliographystyle{splncs04}
% \bibliographystyle{unsrt}


\bibliography{ref}

\end{document}